\documentclass{aa}
\usepackage{graphicx,fixltx2e}
\usepackage{txfonts,hyperref}

\usepackage{natbib} 
\bibpunct{(}{)}{;}{a}{}{,}

\begin{document}

 \title{The energy of waves in the photosphere and lower
   chromosphere: III.~Inversion setup for \ion{Ca}{ii} H spectra in local thermal equilibrium}

   \author{C. Beck\inst{1} \and R. Rezaei\inst{2} \and K.G. Puschmann\inst{3}}
        
   \titlerunning{The energy of waves: III.~LTE inversion of \ion{Ca}{ii} H spectra}
  \authorrunning{C. Beck, R. Rezaei, K.G. Puschmann}  
\institute{Instituto de Astrof\'isica de Canarias (IAC)
         \and Kiepenheuer-Institut f\"ur Sonnenphysik (KIS)
        \and Leibniz-Institut f\"ur Astrophysik Potsdam (AIP)
       }
 
\date{Received xxx; accepted xxx}
\abstract{The \ion{Ca}{ii} H line is one of the strongest lines in the solar
  spectrum and provides continuous information on the solar atmosphere from
  the photosphere to the lower chromosphere.}{We describe an inversion approach that reproduces observed \ion{Ca}{ii} H spectra assuming local thermal equilibrium (LTE).}{We developed an inversion strategy based on the SIR code that reproduces \ion{Ca}{ii} H spectra in the LTE approximation. The approach uses a two-step procedure with an archive of pre-calculated spectra to fit the line core and a subsequent iterative modification to improve the fit mainly in the line wing. Simultaneous spectra in the 630\,nm range can optionally be used to fix the continuum temperature. The method retrieves one-dimensional (1D) temperature stratifications neglecting lateral radiative transport. Line-of-sight velocities are included post-facto by an empirical approach.}{An archive of about 300.000 pre-calculated spectra is more than sufficient to reproduce the line core of observed \ion{Ca}{ii} H spectra both in quiet Sun and in active regions. The subsequent iterative adjustment of the thermodynamical stratification matches observed and best-fit spectra to a level of about 0.5\,\% of $I_c$ in the line wing and about 1\,\% of $I_c$ in the line core.}{The successful application of the LTE inversion strategy suggests that inversion schemes based on pre-calculated spectra allow one a reliable and relatively fast retrieval of solar properties from observed chromospheric spectra. The approach can be easily extended to an 1D non-LTE (NLTE) case by a simple exchange of the pre-calculated archive spectra. Using synthetic NLTE spectra from numerical three-dimensional (3D) simulations instead will finally allow one to extend the approach from the static 1D-case to dynamical atmosphere models including the complete 3D radiative transport.} 
\maketitle
\section{Introduction}
{\em ``The properties of the solar chromospheric plasma are so poorly known that all careful studies would be enlightening, no matter how crude.''} \citep{cram1977}

The spectral lines of \ion{Ca}{ii} H at 396.8\,nm and \ion{Ca}{ii} K at
393.3\,nm are among the strongest and deepest spectral lines in the visible solar spectrum. The intensity radiated away inside these lines comes from layers between the continuum forming layer and the lower chromosphere, spanning a
range of about 1-2\,Mm of height in the solar atmosphere \citep{vernazza+etal1981,carlsson+stein1997}. Both lines have thus
been used intensively for studies of the chromosphere, where useful spectral lines are scarce. Ground-based observations are mainly limited to \ion{Ca}{ii} H and K \citep[e.g.,][]{jensen+orrall1963,beckers1968,liu1974,teplitskaya+firstova1976,cram+dame1983,rutten+uitenbroek1991,lites+etal1999,rezaei+etal2007,beck+etal2008,teplitskaya+etal2009}, H$\alpha$ \citep[e.g.,][]{bray1973,schmieder+etal1984,tsiropoula+etal1993,balasubramaniam+etal2004,lopezariste+etal2005,sanchezandrade2008,cauzzi+etal2009,bostanci2011}, the \ion{Ca}{ii} IR triplet \citep[e.g.,][]{lopezariste+etal2001,socasnavarro2005,pietarila+etal2007,cauzzi+etal2008,vecchio+etal2008}, and \ion{He}{i} at 1083\,nm \citep{lites1986,ruedi+etal1995,socasnavarro+elmore2005,sanchezandrade+etal2007,kuckein+etal2009,felipe+etal2010} by the absorption in the Earth's atmosphere, whereas from space also \ion{Mg}{ii} h and k, L$\alpha$, and other lines in the extreme UV are accessible \citep[e.g.,][]{kneer+etal1981,bonnet1981,fontenla+etal1988,carlsson+etal1997,curdt+etal2010}\footnote{The references here and before are only intended to cover roughly every decade and research group/instrument.}. 

Despite the fact that the \ion{Ca}{ii} H and K lines provide a tremendous amount of information on all of the lower solar atmosphere, it turned out to be rather difficult to extract this information from observed spectra \citep{linsky+avrett1970}. Part of the spectral lines forms in an atmospheric regime where the gas density is so low that the assumption of instantaneous local thermal equilibrium (LTE) is no longer valid, because collisions are so scarce that the energy is not necessarily distributed evenly over all possible degrees of freedom. A consistent treatment of line formation in non-LTE (NLTE) is, however, at the limits of the theoretical description and the commonly available computational resources \citep{solanki+etal1991,carlsson+stein2002,leenaarts+etal2009,martinezsykora+etal2012}. After a few attempts for a direct analysis of \ion{Ca}{ii} spectra in quiet Sun (QS) \citep[][]{liu+skumanich1974} or active regions \citep[e.g.,][]{teplitskaja+efendieva1975} in the 70's and 80's, the topic has not received much attention in the recent years. Apart from the inversion code for \ion{Ca}{ii} K spectra described in \citet{rouppevandervoort2002} that is based on \citet{shine+linsky1974a} and the manual inversion method for the line wings of \ion{Ca}{ii} H and K used in \citet{sheminova+etal2005} and \citet{sheminova2012}, only the group led by Prof.~Teplitskaya seems to have continued with the development of analysis and/or inversion tools for \ion{Ca}{ii} H and K \citep{grigoryeva+etal1991,grigoryeva+etal2000,grigoryeva+etal2009a,teplistkaya+grigoryeva2009}. 

Many other studies dealt with \ion{Ca}{ii} H and K only in the forward modeling direction, i.e., creating spectra from a model atmosphere to match some average characteristic profile \citep[e.g.,][]{linsky+avrett1970,shine+linsky1974,suemoto1977,kneer+mattig1978,lites+skumanich1982,solanki+etal1991}, but not with the intention to analyze individual spectra in a two-dimensional (2D) field of view (FOV). The formation of \ion{Ca}{ii} H spectra was addressed in numerical NLTE simulations \citep[e.g.,][]{rammacher+ulmschneider1992,rammacher+cuntz2005,carlsson+stein1997}, but again in the forward modeling approach, i.e., the synthesis of \ion{Ca}{ii} spectra from the output of a numerical model \citep[see also][]{uitenbroek2011}. While this provides consistent NLTE spectra, it does not allow one to analyze observed spectra directly. 

For photospheric spectra, several inversion codes exist that allow one to retrieve physical parameters of the solar atmosphere from observed spectra on an optical depth scale \citep[e.g.,][]{cobo+toroiniesta1992,socas+etal2001}. With (a few) additional assumptions, the inversion results can then also be converted to an absolute geometrical height scale \citep[e.g.,][]{sanchezalmeida2000,puschmann+etal2005,puschmann+etal2010,beck2011}. The geometrical height scale is essential for determining many quantities derived from the magnetic field topology such as the electric current density or the helicity in sunspots \citep{puschmann+etal2010a,cobo+puschmann2012}. For observations of chromospheric spectral lines, similar inversion codes are only partially available \citep{socasnavarro+etal1998,tziotziou2007}. For a possible application to \ion{Ca}{ii} H spectra, the only new development is the NLTE inversion code NICOLE based on \citet{socas-navarro+etal2000} that was used in, e.g., \citet{socasnavarro+etal2006}, \citet{pietarila+etal2007a} and \citet{delacruz+etal2012}. However, it still needs to be tested if it can be successfully applied to \ion{Ca}{ii} H spectra. 

For the analysis of \ion{Ca}{ii} H spectra one therefore has to rely on methods
that are (hopefully) insensitive to NLTE effects, using observed
quantities as direct as possible to derive solar atmospheric properties \citep[e.g.,][]{beck+rammacher2010}. In
the first two papers of this series \citep[][A\&A, in press; BE09 and BE12 in the following]{beck+etal2009,beck+etal2012}, we investigated the statistics of velocity and intensity oscillations in \ion{Ca}{ii} spectra, using only \ion{Ca}{ii} H for the first study, and also \ion{Ca}{ii} IR at 854\,nm for the latter thanks to new observations covering both lines simultaneously. The conversion from observed intensities or velocities to energy in these two studies necessarily had to rely on some
assumptions such as the LTE condition that have an impact on the final result. A consistent NLTE analysis would be strongly preferred, but lacking a suitable method for the automatic treatment of the observations with several 10.000 spectra each, we now first developed and applied an LTE inversion strategy for \ion{Ca}{ii} spectra based on the SIR code \citep{cobo+toroiniesta1992} as preliminary step to a possible full NLTE treatment in the future. The LTE assumption does not fail instantly in the solar atmosphere, but results derived assuming LTE will deviate more and more from the ``reality'' the closer one gets towards the very line core of \ion{Ca}{ii} H that forms highest above the continuum. The LTE inversion can also be helpful for an NLTE analysis: it both outlines a suitable approach for the inversion procedure and provides a reasonable initial model for the NLTE fit that presumably will speed up its convergence.  

The observations used are outlined briefly in Sect.~\ref{observations}, whereas Sect.~\ref{stray} describes the stray-light correction. The LTE inversion approach is described in Sect.~\ref{inv_code}. Section \ref{results} shows the result of the application of the code to the spectra. The results are discussed in Sect.~\ref{discussion}. Section \ref{conclusion} provides our conclusions. Appendix \ref{appa} explains the animation of the inversion results that accompanies this paper.
\section{Observations\label{observations}}
For the present study, we used the \ion{Ca}{ii} H spectra of the QS observations on disc centre labeled No.~1 and 2 in BE12. The first observation is a large-area scan taken on 24 July 2006 that was used in BE09 before and is described in detail there. It consists of simultaneous spectroscopy of \ion{Ca}{ii} H and spectropolarimetry at 630\,nm obtained with the POlarimetric LIttrow Spectrograph \citep[POLIS,][]{beck+etal2005b} at the German Vacuum Tower Telescope \citep[VTT,][]{schroeter+soltau+wiehr1985} in Iza{\~n}a, Tenerife, Spain. The integration time per scan step was 6.6\,secs. The image of the solar surface was scanned in 150 steps of 0\farcs5 step width, covering a total FOV of about 75$^{\prime\prime} \times 70^{\prime\prime}$. The slit width was 0\farcs5, and the spatial sampling along the slit was 0\farcs3. The second observation is a time-series taken with POLIS on the same day that is described in detail in \citet[][BE08]{beck+etal2008}. The integration time was 3.3\,secs per scan step and the cadence of repeated co-spatial spectra is about 21\,secs. 

For comparison with the results for these QS data on disc centre we also applied the inversion code to an observation of an active region (AR) off the disc centre at a heliocentric angle of about 50$^\circ$. These data were taken on 8 December 2007 with a combination of POLIS and the Tenerife Infrared Polarimeter \citep[TIP,][]{martinez+etal1999,collados+etal2007}. The data consist of three scans of 200 steps with a step width of 0\farcs5, covering a great part of the active region NOAA 10978. The integration time was 3.3\,secs per scan step. The rest of instrumental characteristics for POLIS was as above. Overview maps of these observations can be found in \citet[][]{beck+rammacher2010}, their Fig.~16. One of the AR maps is described in detail in \citet[][]{bethge+etal2012}. All data were taken with real-time seeing correction by the Kiepenheuer-Institute adaptive optics system \citep[KAOS,][]{vdluehe+etal2003}. The spatial resolution of the observations, being slit-spectrograph data, cannot be improved post-facto by image reconstruction techniques \citep[e.g.,][]{puschmann+beck2011} apart from a possible correction for the instrumental point spread function \citep[][BE11 in the following]{beck+etal2011}.
\section{Stray-light correction\label{stray}}
\begin{figure}
\includegraphics{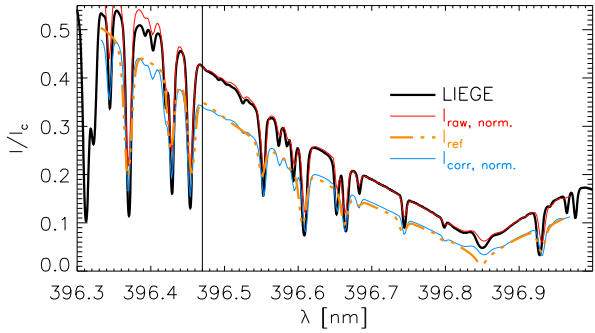}
\caption{Intensity normalization and stray-light correction of the average \ion{Ca}{ii} H profile. {\em Thick black}: LIEGE atlas profile. {\em Thin red}: normalized average observed profile $I_{\rm raw, norm.}$. {\em Thick orange dash-dotted}: reference profile of the modified HSRA model. {\em Thin blue}: stray-light corrected average observed profile $I_{\rm corr., norm.}$. The {\em black vertical line} denotes the wavelength of intensity normalization.\label{fig_int_norm}}
\end{figure}

The \ion{Ca}{ii} H channel of POLIS has two major contributions of stray-light (BE11): a spectrally undispersed (``parasitic'') part, $\beta$, caused by scattering inside of the spectrograph itself, and spectrally resolved stray-light, $\alpha$, caused by optics inside and upfront of POLIS. For the current investigation, we used a simplified version of the stray-light correction described in BE11. The observed spectra after subtraction of the dark current and flat fielding, $I_{\rm raw}$, were first corrected for the prefilter transmission curve and normalized on average to the Liege spectral atlas \citep{delbouille+etal1973} in the line wing near 396.5\,nm ({\em vertical black line} in Fig.~\ref{fig_int_norm}). The normalized spectra $I_{\rm raw, norm.}$ were then corrected for stray-light by
\begin{equation}
I_{\rm corr., norm.} = (I_{\rm raw, norm.} - 0.025)\cdot 0.85 \,. \label{eq1}
\end{equation}
This corresponds to assuming a parasitic stray-light level of $\beta= 2.5\,\%$ of $I_c$ and a spectrally resolved stray-light level of $\alpha=15\,\%$. The former value is identical to that derived in BE11 (5\,\% of the intensity at 396.4\,nm $\sim$ 2.5\,\% of $I_c$), whereas $\alpha$ is slightly lower. 

The values of $\alpha$ and $\beta$ used here were determined to match the average observed profile and a reference profile related to the inversion of the spectra. The reference profile ({\em dash-dotted orange line} in Fig.~\ref{fig_int_norm}) corresponds to the LTE synthesis of the \ion{Ca}{ii} H spectrum for a model atmosphere without a chromospheric temperature rise constructed from a combination of the Harvard Smithsonian Reference Atmosphere \citep[HSRA,][]{gingerich+etal1971} and the Holweger-Mueller model \citep[HOLMUL,][]{holweger+mueller1974}. The exact shape of the reference profile in the line-core, i.e., from 396.83\,nm to 396.88\,nm, where the emission peaks and the central absorption core are located, is uncritical because only the line wings are matched by the stray-light correction, not the very line core. Therefore the choice of which temperature stratification to use for creating the reference profile \citep[HOLMUL, HSRA, VAL, FAL;][]{vernazza+etal1981,fontenla+etal2006} is not important because all these theoretical atmosphere models only differ significantly in chromospheric layers, and hence in the shape of the line core, whereas the line wing is similar for spectra synthesized from any of the models. 
\section{LTE inversion\label{inv_code}}
Initial attempts to fit the Ca spectra with the SIR code (BE08, Appendix B) showed that the code tends to ignore the line-core region in the fit. This is caused by the equal weight used for every wavelength point, which is perfectly suitable for photospheric lines, but not for chromospheric ones. The line core and the neighboring wavelength points ($\pm 0.05$\,nm $\equiv \pm$ 25 pixels of 1.92 pm in the POLIS spectra) contain all the information about the ``upper'' part of the atmosphere ($\log \tau < -2$), whereas the remaining about 270 wavelength points contain (redundant) information about the lower atmosphere. The intensity near the line core is the lowest of the full spectrum, and therefore any mismatches between synthetic and observed spectra stay small and have no strong effect on the least-square value to be minimized in the fit.

A second problem is that the temperature stratifications required to reproduce the observed spectra may have very complicated shapes such that an iterative modification of a given simple initial model may never result in the necessary complex shape, regardless of the number of nodes used for modifying the temperature stratification during the fit. To overcome this limitation in the inversion of the \ion{Ca}{ii} H spectra, we used as first step a comparison of the observed spectra to a pre-calculated archive of \ion{Ca}{ii} H spectra, giving an increased weight to the line core at this step. Similar archives have been used in inversion approaches of other chromospheric lines before, mainly for H$\alpha$ \citep{molownyhoras+etal1999,tziotziou+etal2001,schmieder+etal2003,berlicki+etal2005}. 
\subsection{Creation of an archive of LTE spectra\label{create_arch}}
The base of the archive is the modified version of the HSRA that emulates an atmosphere in radiative equilibrium. The archive profiles were then generated by adding temperature perturbations of varying shapes and amplitudes to the modified HSRA model and synthesizing the resulting spectra with the SIR code. SIR uses complete frequency redistribution and assumes LTE conditions. We only modified the temperature and kept all other parameters as given in the modified HSRA model. 

\begin{figure}
\centerline{\resizebox{8.8cm}{!}{\includegraphics{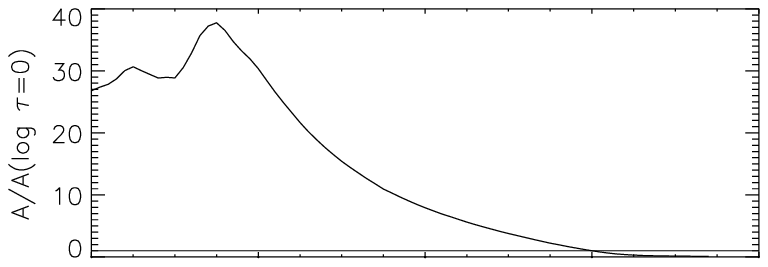}}}\vspace*{-1.35cm}
\centerline{\resizebox{8.8cm}{!}{\includegraphics{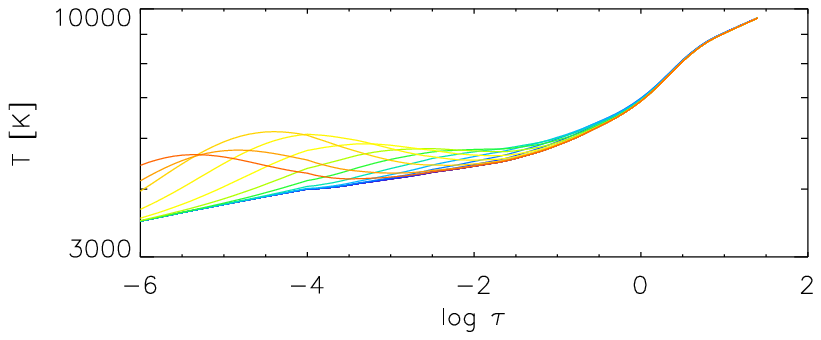}}}
\centerline{\hspace*{1.cm}\resizebox{6.8cm}{!}{\includegraphics{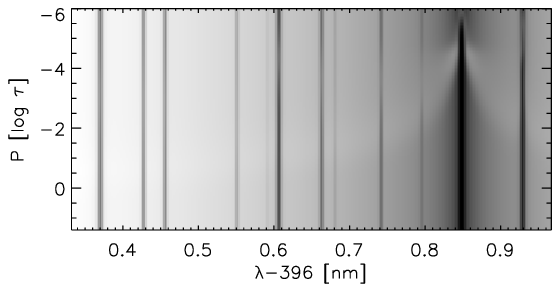}}}$ $\\
\caption{{\em Top}: scaling law for the temperature amplitude with optical depth. {\em Middle}: example of temperature stratifications with the Gaussian perturbation centred subsequently at decreasing optical depth. {\em Bottom}: the 75 synthesized spectra corresponding to the temperature stratifications above for the Gaussian perturbation centred subsequently at the location $P$ from +1.4 ({\em bottommost} spectrum) to $-$6 ({\em topmost} spectrum) in log $\tau$. \label{fig_temp}}
\end{figure}

One example of the temperature perturbations applied in creating the archive was of Gaussian shape, with different central positions $P$ in log $\tau$, widths $\sigma$, and amplitudes $T_{ampl}$ (at log $\tau$=0). The ranges for the three parameters are given in Table \ref{tab_archive}. The temperature amplitude for the Gaussian was scaled up with its location in optical depth by the square-root of the electron pressure, normalized to its value at log $\tau$=0 (cf.~the {\em top panel} of Fig.~\ref{fig_temp}). The Gaussian perturbation was then moved across all 75 grid points in optical depth from log $\tau$=$+$1.4 to $-$6. An example of the resulting temperature stratifications for one run of the Gaussian through optical depth is given in the {\em middle panel} of Fig.~\ref{fig_temp}. The {\em lower panel} shows the resulting synthetic spectra. This setup was chosen because spectra from a single run in optical depth roughly reproduce the temporal evolution of observed spectra during the occurrence of a bright grain (see the {\em right panel} of Fig.~\ref{slit_spec} later on). Note that because of the scaling of the temperature amplitude with optical depth, an initial amplitude $T_0$ at log $\tau$=0 of, e.g., 100\,K converts to a perturbation with an amplitude of about 4000\,K when the Gaussian is located at $\log\,\tau \sim -4$. The archive thus automatically covers a large range of possible temperature values. We additionally globally added (subtracted) up to $\pm 300$ K to (from) all depth points ($T_{diff}$, {\em first column} of Table \ref{tab_archive}) when running the Gaussian perturbation through optical depth. Additional synthetic spectra were created by adding (subtracting) straight lines of variable slope to (from) the modified HSRA model instead of adding a Gaussian perturbation. 

Line-of-sight (LOS) velocities were not included in the generation of the archive profiles because this would have increased the size of the archive beyond
usefulness. Velocities are instead taken into account in an empirical way in a later step. The spectral resolution of the \ion{Ca}{ii} H spectra from POLIS corresponds to a velocity dispersion of about 1.5\,kms$^{-1}$ per pixel. Together with the intrinsic (thermal) width of spectral features, the weak velocity response allowed us to still use the pre-calculated archive spectra as first step to fix the temperature stratification because the Doppler shifts (in pixel) stay comparably small. The macroturbulent velocity in the atmosphere was set to a fixed value of 1\,kms$^{-1}$ in the generation of the archive spectra, which yielded line widths for the photospheric line blends that match the observed line widths.
\begin{table}
\caption{Parameter ranges used in generating the archive. The values in parentheses give the step width.\label{tab_archive}}
\begin{tabular}{cccc}
$T_{diff} [K]$ & $T_{ampl}|_{\tau = 1}$ [K] & $\sigma$ [log$\tau$] & $P$ [log$\tau$]\cr\hline
-300  - +300 (50)  & 10 - 100  (10)  & 0.3 - 3.1 (0.1) & -6 - +1.4 (0.1)\cr
 \end{tabular}
\end{table}
\begin{figure*}
\begin{minipage}{12cm}
\resizebox{12cm}{!}{\includegraphics{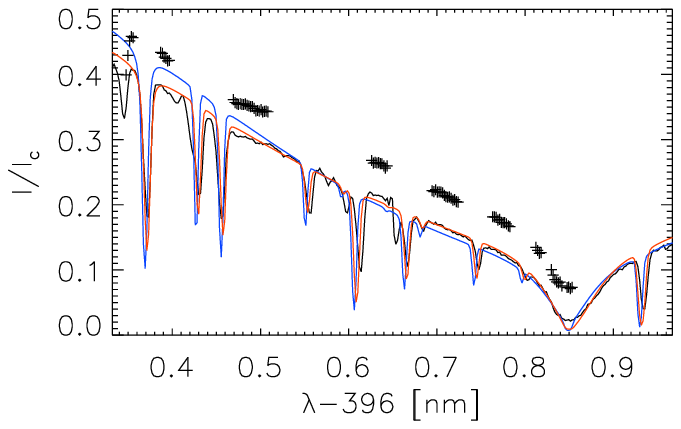}\includegraphics{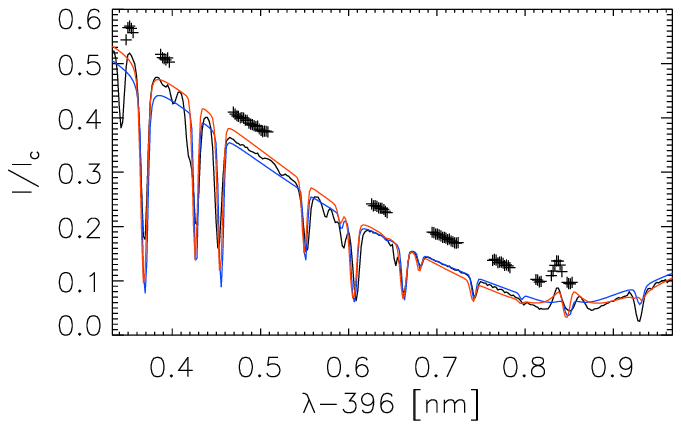}}\vspace*{-.4cm}\\
\resizebox{12cm}{!}{\includegraphics{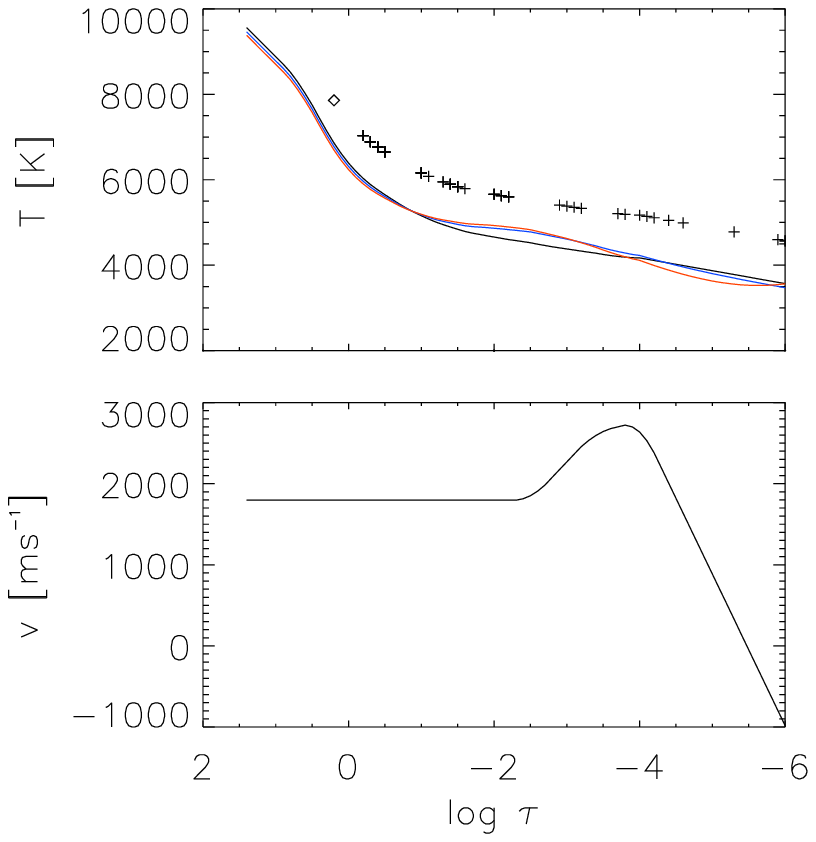}\includegraphics{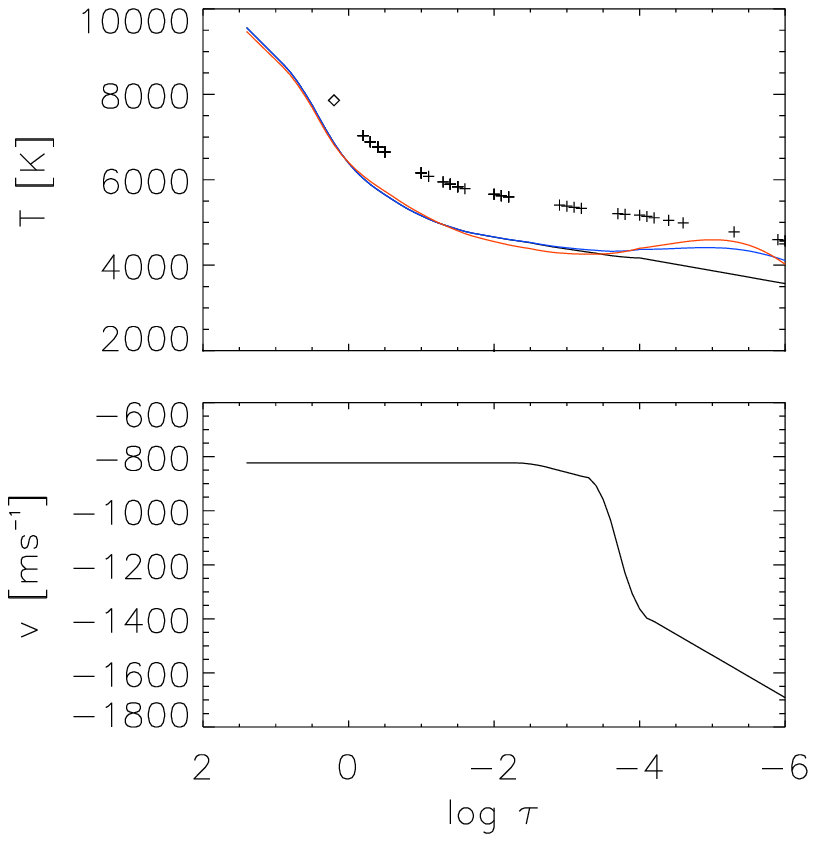}}\\
\end{minipage}
\begin{minipage}{5.6cm}
\caption{Iterative improvement of the best-fit archive spectrum for two example spectra without ({\em left column}) and with emission in the line core ({\em right column}). {\em Top row}: observed spectrum ({\em black line}), best-fit archive spectrum ({\em blue}), and final best-fit spectrum ({\em red}). The {\em black crosses} denote the wavelengths and intensities used for the improvement of the fit. {\em 2nd row}: temperature stratifications ({\em blue}: archive best-fit, {\em red}: final best-fit, {\em black}: modified HSRA model). The {\em black crosses} denote the optical depths corresponding to the wavelengths marked above. The {\em diamond} at log $\tau =$ 0.2 denotes the 630 nm continuum intensity contribution. {\em Bottom row}: stratification of the LOS velocity.\label{fit_iter}}  
\end{minipage}
\end{figure*}
 
With the Gaussian perturbations defined by the free parameters ($T_{diff}$, $P$, $\sigma$, $T_{ampl}$) and the other additional modifications of the temperature stratification in total more than 300.000 \ion{Ca}{ii} H spectra were synthesized with SIR. All line blends of \ion{Ca}{ii} H with known transition parameters were taken into account (cf.~Table 1 in BE09). We also always synthesized from the atmospheric models the corresponding spectra at 630\,nm because POLIS provides both wavelength ranges simultaneously. The creation of the archive by a single (background) process takes about one week.
\subsection{Best-fit spectrum from LTE archive}
To find the archive profile $A(\lambda,i)$ ($i$ counts the
number inside the archive) that matches best a given observed profile
$I(\lambda)$, the least-square deviation between the observed profile
and all archive profiles was calculated by
\begin{equation}
\chi^2_i = \sum_j w(\lambda_j) \cdot ( I(\lambda_j) - A(\lambda_j,i) ) ^2 \;. \label{chieq}
\end{equation}
The weights $w(\lambda_j)$ were set to the squared inverse of an average QS profile, $w = 1/I_{\rm av}^2$, with a subsequently doubled weight for the region around the line core itself from about 396.81\,nm to 396.89\,nm. This strongly enhances the influence of the line-core intensities relative to the wing (ratio of about 4:1). All (strong) line blends were masked out by setting the weight to zero at these wavelengths. The 630\,nm spectra were not included in the $\chi^2$ for finding the best-matching archive profile because the fit to the Ca line core is the main driver for this step. 
\subsection{Iterative improvement of temperature stratification\label{mod_temp}}
The best match of observed and archive profile is determined with a strong
weight for the very line core. In the line wing, the observed and
 best-fit archive profile usually deviate because the archive only contains
a limited number of all of the possible combinations of relative wing and core
intensities. To improve the fit in the line wing, we used an iterative modification of the temperature stratification based on the intensity differences between observed and synthetic profiles at some wavelengths. The temperature values at the corresponding layers in log $\tau$ that contribute to the intensity at these wavelengths are then modified according to the modulus and sign of the intensity differences.

We selected a series of wavelength windows without photospheric blends and the blue part of the line core up to the line centre (cf.~the {\em black crosses} in the {\em top row} of Fig.~\ref{fit_iter}). For these wavelengths $\lambda$, we calculated the corresponding layers in log $\tau$ using the intensity response function of \ion{Ca}{ii} H in the modified HSRA model \citep[cf.][BE09]{rezaei+etal2008}. The intensity response function matches well formation heights derived from phase differences of propagating waves (BE08, BE09). We then implemented the relation between $\lambda$ and corresponding $\tau$ into the code (provided by an external file). It turned out that the hardwired response function works fine for all spectra on disc centre, but for the AR observations at a heliocentric angle of 50$^\circ$, the relation between $\lambda$ and $\tau$ was already slightly off leading to a misfit in the outermost wing/lowermost atmosphere layers. It seems therefore to be recommended to at least re-calculate the average $\lambda$-$\tau$ relation for an application to off-centre observations. 

During the iterative improvement of the complete temperature stratification also the simultaneous 630\,nm spectra were now included by adding one more point to the wavelength windows used. This point was located in the continuum of the 630\,nm channel, corresponding to $\log\tau \sim 0.2$ in the intensity response. The lowermost forming wavelengths in the Ca wing in the usual setup of POLIS form at about $\log\tau \sim -0.2$ (cf.~the {\em black crosses} in the {\em middle row} of Fig.~\ref{fit_iter}) The use of the 630\,nm continuum intensity was, however, implemented only optionally and can be skipped when no 630\,nm data are available.  

With the relation between $\lambda$ and $\tau$ given, one  can then modify the
temperature stratification according to the intensity differences between
observed and best-fit spectrum, $\Delta I(\lambda)$, at the wavelengths
considered. The intensity difference is first converted to a temperature difference using the intensity response function. We then fit a fourth-order polynomial to the required temperature changes in $\log \tau$ to obtain a smooth curve for all optical depth values. The slope of the curve is extrapolated towards the layers with $\log\tau>0.2$ where no measurements are available. The change of the
temperature stratification is then added to the previous stratification, and
the new spectrum is synthesized. The {\em middle panels} of
Fig.~\ref{fit_iter} show some examples of the variations during the iteration
of the temperature stratification. The method converges usually on two, three iterations. The code uses no special convergence criterion, only the number of iterations to be done has to be provided by the user.

In the iteration, no special weight is given to the line core because this step is mainly intended to improve the fit in the line wing. As a result, the $\chi^2$ value using the weighting with an enhanced contribution of the line core (cf.~the previous section) can actually worsen. Even if this happens, the {\em iterated} temperature stratification is returned because otherwise it is not possible to obtain a good fit to the line wing. To identify such cases, i.e., a degradation of the fit to the Ca line core in the iteration process, the inversion routine returns an additional binary value of 0 or 1 that indicates a decrease or increase of the weighted $\chi^2$ (cf.~Sect.~\ref{qual_fit} later on).

The LTE condition in the spectral synthesis of the profiles requires the temperature stratifications to have a negative slope in the uppermost layers for creating double reversals in the line core (cf.~the {\em middle right panel} of Fig.~\ref{fit_iter}). This usually affects the last 3\,--\,5 grid points of the optical depth scale ($\log \tau = -5.5\,$ to $-6$) at the upper end of the optical depth scale. The profiles created by the addition of the Gaussian perturbation to the modified HSRA automatically have such a negative slope whenever the centre of the Gaussian is not located exactly at $\log \tau = -6$.
\begin{figure*}
\begin{minipage}{11cm}
\centerline{\resizebox{11.cm}{!}{\hspace*{.5cm}\includegraphics{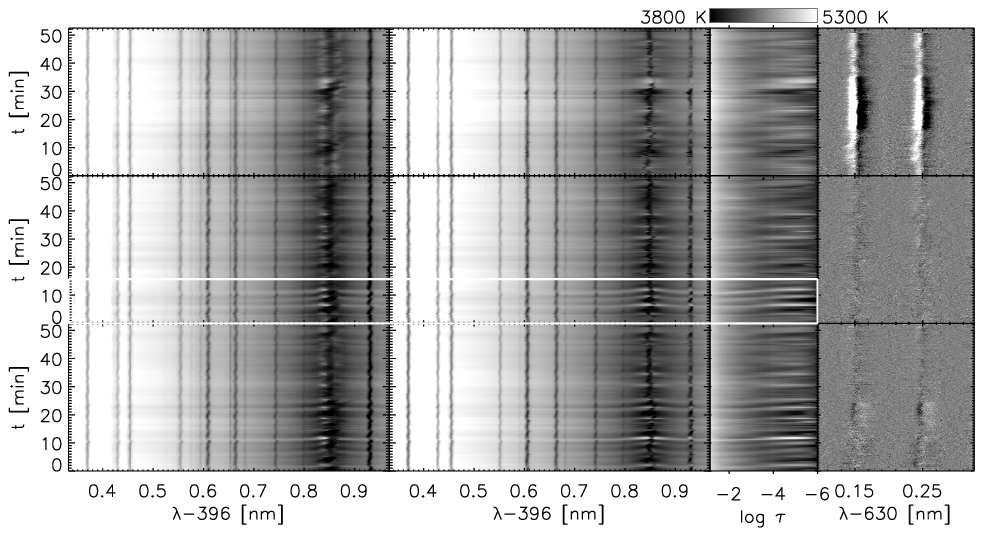}}}$ $\\
\end{minipage}
\begin{minipage}{6.6cm}
\vspace*{.3cm}
\resizebox{6.2cm}{!}{\hspace*{1.25cm}\includegraphics{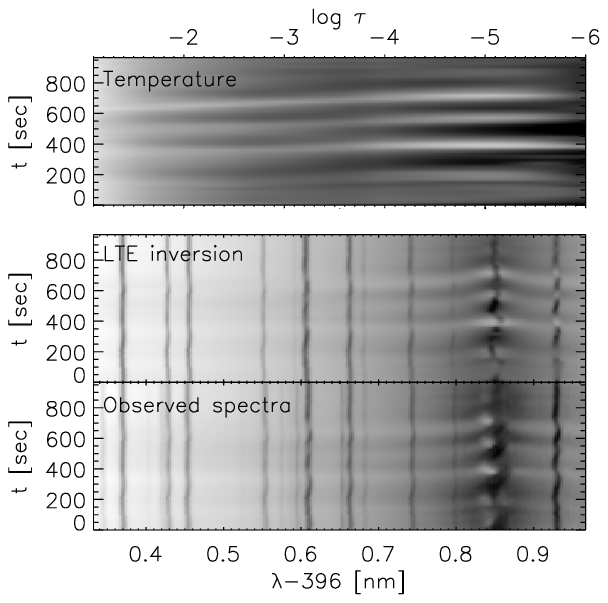}}\\$ $\\
\end{minipage}
\caption{Comparison of observed spectra and inversion results in the time-series. {\em Left panel, top to bottom}: spectra from three different locations along the slit versus (vs.) time. The {\em upper row} corresponds to a network location. {\em First column}: observed spectra. {\em Second column}: best-fit spectra. The {\em third column} shows the temperature stratifications vs.~$\log \tau$, the {\em fourth column} the simultaneous Stokes $V$ spectra of the 630 nm channel. The {\em white rectangle} in the {\em second row} marks the section shown magnified at the {\em right-hand side}. A train of three consecutive brightenings passes from the line wing towards the line core between $t$\,=\,300\,secs and $t$\,=\,800\,secs.\label{slit_spec}}
\end{figure*}
\begin{figure*}
\begin{minipage}{11cm}
\centerline{\resizebox{11.cm}{!}{\hspace*{.5cm}\includegraphics{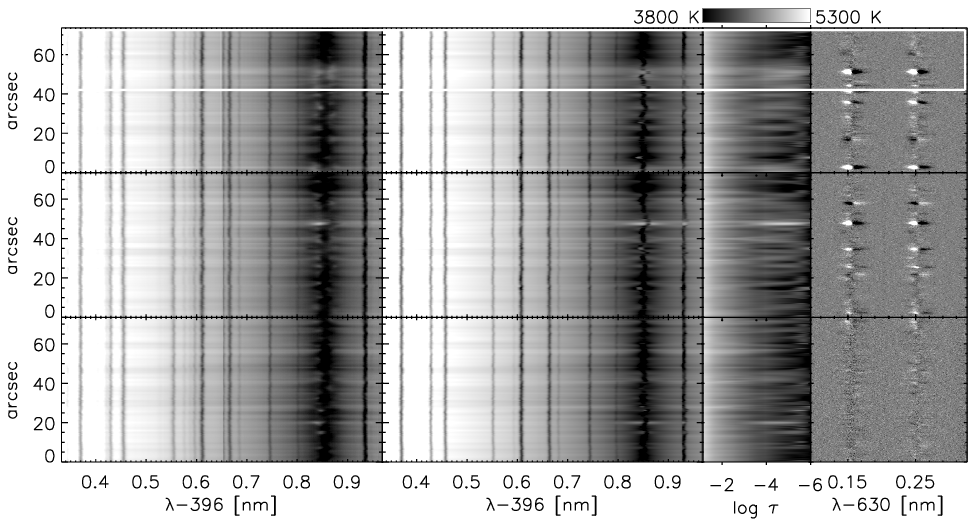}}}$ $\\
\end{minipage}
\begin{minipage}{6.6cm}
\vspace*{.0cm}
\resizebox{4.9cm}{!}{\hspace*{1.25cm}\includegraphics{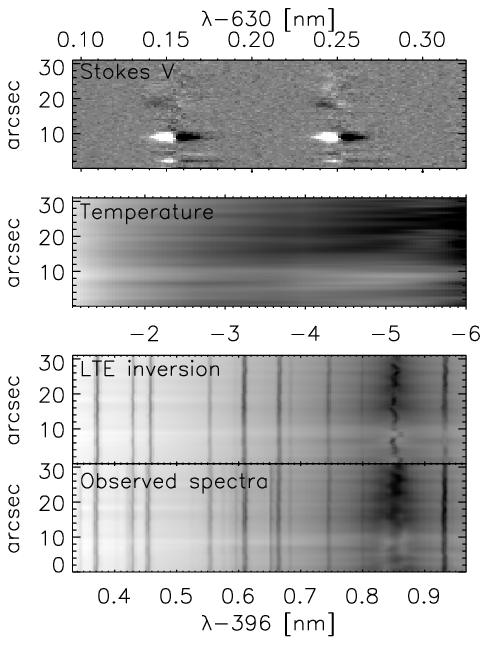}}\\$ $\\
\end{minipage}
\caption{Same as Fig.~\ref{slit_spec} for the large-area scan. Note that this time the cuts are along a spatial axis and do not represent a temporal evolution. The {\em white rectangle} in the {\em top row} marks the section shown magnified at the {\em right-hand side}.\label{spat_spec}}
\end{figure*}
\begin{figure*}
\sidecaption
\begin{minipage}{12cm}
\centerline{\resizebox{12.cm}{!}{\hspace*{.75cm}\includegraphics{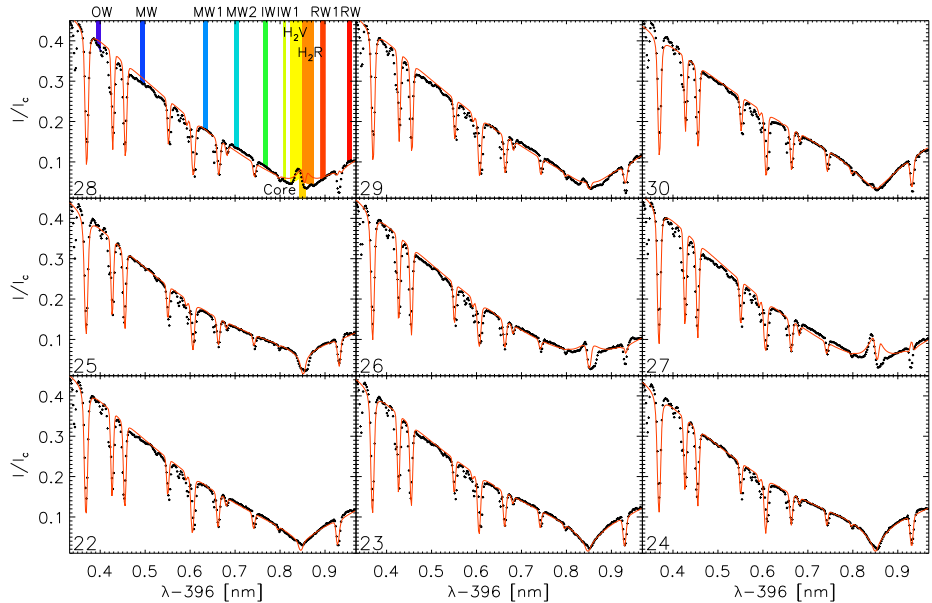}}}$ $\\
\end{minipage}
\caption{Example spectra during the passage of a (shock) wave. Observations are given by {\em black pluses} and the best-fit spectra by {\em red lines}. The temporal cadence between subsequent spectra is 21 sec. Time increases from {\em left to right} in each row and from {\em bottom to top} between rows. The {\em coloured bars} in the {\em upper left panel} denoted the wavelength bands used later on.\label{fig5} } 
\end{figure*}
\subsection{Inclusion of LOS velocities}
In the iterative improvement of the temperature stratification, the LOS velocities are finally included in an empirical way. We use three of the line blends of \ion{Ca}{ii} H (\ion{Cr}{i} at 396.37\,nm, \ion{Fe}{i} at 396.61\,nm, \ion{Fe}{i} at 396.93\,nm) and the Ca line core itself. The line blends in the wing correspond to three different formation heights (cf.~BE09). We use the observed LOS line-core velocities of the four lines to construct a velocity stratification by attributing the observed velocities to the corresponding $\log\tau$ layers of $-2.4, -3.4, -3.9$, and $-6$, and then linearly interpolate between the velocity values. To prevent step functions, the resulting velocity curve with optical depth is smoothed. The {\em bottom panels} of Fig.~\ref{fit_iter} show the resulting velocity stratification for the observed spectra in the {\em top panels}. The velocity is not modified to obtain a better fit, it serves only to yield a better match of synthetic and observed spectra using the observed velocity values. For the line blends, the velocity is well determined, whereas for the Ca line core the location of the line minimum in some cases might not reflect a true velocity at all.

For the inversions of the AR maps, we did not include LOS velocities because the line core of \ion{Ca}{ii} H has there often a complex shape with a single (or several) intensity reversals that does not easily allow one to ascribe any velocity to it. Because of the weak response of the spectra to LOS velocities, this should have had a negligible effect on the retrieved temperatures. 
\begin{figure*}
\sidecaption
\begin{minipage}{12cm}
\centerline{\resizebox{11.5cm}{!}{\hspace*{.5cm}\includegraphics{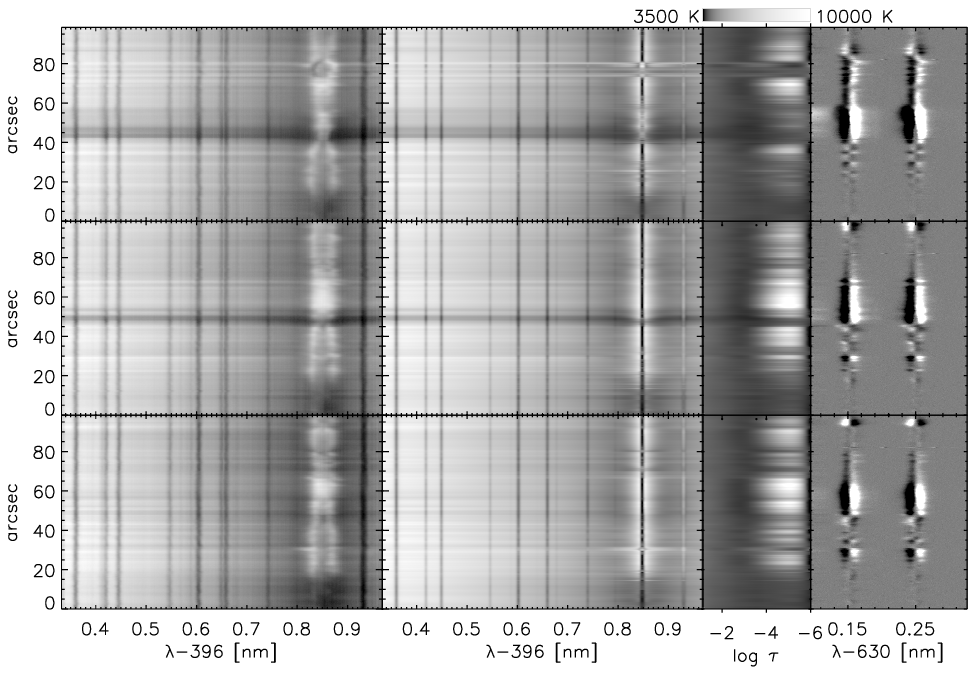}}}$ $\\
\end{minipage}
\caption{Same as Fig.~\ref{spat_spec} for three cuts through the active region. No LOS velocities were considered in this fit. The temperature is displayed on a logarithmic scale in a range from 3500\,K to 10000\,K this time.\label{spat_spec_ar}}
\end{figure*}

It takes about 8 seconds to obtain a fit to a single observed profile on a common desktop computer, and hence about three days for all observed spectra in one typical observation of 30.000 profiles.
\section{Results\label{results}}
\subsection{Example spectra in quiet Sun\label{qs_spec_exam}}
For a visual inspection of the quality of the fits, we show a series of spectra in the following Figs.~\ref{slit_spec} to \ref{fig_ar_single}. The {\em left panel} of Fig.~\ref{slit_spec} shows the observed and best-fit Ca spectra ({\em first} and {\em second column}) for three locations along the slit in the time-series in QS. The {\em third column} shows the corresponding temperature stratifications. The {\em fourth column} shows the simultaneous and co-spatial Stokes $V$ spectra at 630\,nm to demonstrate the presence (or absence) of photospheric magnetic fields. The {\em upper row} corresponds to a location in or near the photospheric network \citep[see also][]{beck+etal2008a}. An animation of the spectra along the slit and the corresponding inversion results for all time steps is available in the online material (see Appendix \ref{appa}). 

Comparing observed and best-fit spectra in Figs.~\ref{slit_spec} to \ref{fig5}, we find that the inversion approach is able to deal well with the spectral range from the wing up to about 396.8\,nm, fully reproducing the intensity patterns and the Doppler shifts of the line blends. In the Ca core, the red emission peak is in general always less well reproduced than the blue one. This is caused by the inability to reproduce the line-core region without including NLTE effects and more complex velocity fields than used in the inversion \citep[cf.][]{carlsson+stein1997}. 

The core of the spectral line of \ion{Fe}{i} at 396.93\,nm reverts to emission in the best-fit spectra on several locations, contrary to the observations (e.g., at $t=400, 600, 750\,$secs in the {\em right panel} of Fig.~\ref{slit_spec}). The LTE assumption thus already fails for the core of this line  in some cases. The temporal evolution, e.g., wave trains propagating from the line wing towards the core, and all variations in the line wing are well captured by the inversion. The behaviour of the observed spectra during the passage of three consecutive wave fronts ({\em right panel} of Fig.~\ref{slit_spec}) is well matched except for the asymmetry of the red and blue emission peak. The corresponding temperature stratifications show both the upwards propagation of the wave fronts and the increase of their amplitude in the upper atmosphere.

The differences between magnetic and field-free locations can be seen better in Fig.~\ref{spat_spec} that shows spatial cuts through the large-area map. Whereas the (temporary) enhancements typical for propagating waves often only appear in the upper layers ($\log \tau < -4$) with a reduction of temperature between about $\log\tau = -2$ to $-4$ (cf.~the {\em top panel} at the {\em right-hand side} of Fig.~\ref{slit_spec}), the temperature on most locations with significant polarisation signal, and hence magnetic fields is usually increased over the complete optical depth range ({\em third column} of the {\em left panel} of Fig.~\ref{spat_spec}). The magnification {\em at the right-hand side} of Fig.~\ref{spat_spec} shows that in some cases regions with reduced intensity in the very line core can extend spatially connected over 5\,--\,10$^{\prime\prime}$, reflected by the corresponding reduction of temperature at $y= 20-30^{\prime\prime}$, in contrast to the magnetic location around $y=9^{\prime\prime}$.
\begin{figure}
\resizebox{8.8cm}{!}{\hspace*{1.2cm}\includegraphics{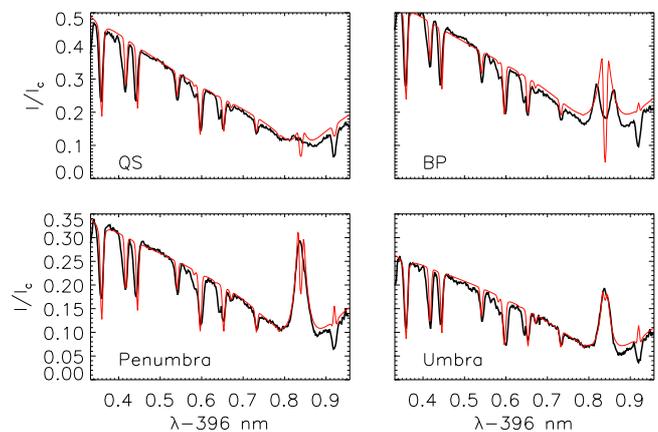}}$ $\\$ $\\
\caption{Individual spectra from the active region map. {\em Clockwise, starting left top}: quiet Sun, bright point in QS, umbra, penumbra. {\em Black}: observed spectrum. {\em Red}: best-fit spectrum.\label{fig_ar_single}}
\end{figure}
\begin{figure*}
\fbox{\begin{minipage}{11.8cm}Temperature\\
\centerline{\resizebox{11.8cm}{!}{\hspace*{.5cm}\includegraphics{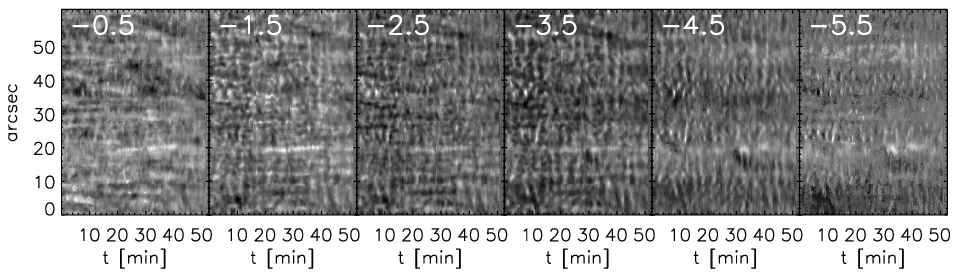}}}$ $\\$ $\\
\centerline{\hspace*{.25cm}\resizebox{8.cm}{!}{\includegraphics{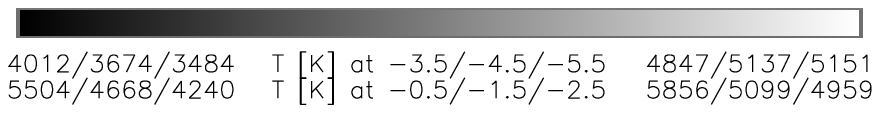}}}
\end{minipage}}$ $\\$ $\\
\fbox{\begin{minipage}{11.8cm}Wavelength bands\\
\resizebox{11.8cm}{!}{\hspace*{.75cm}\includegraphics{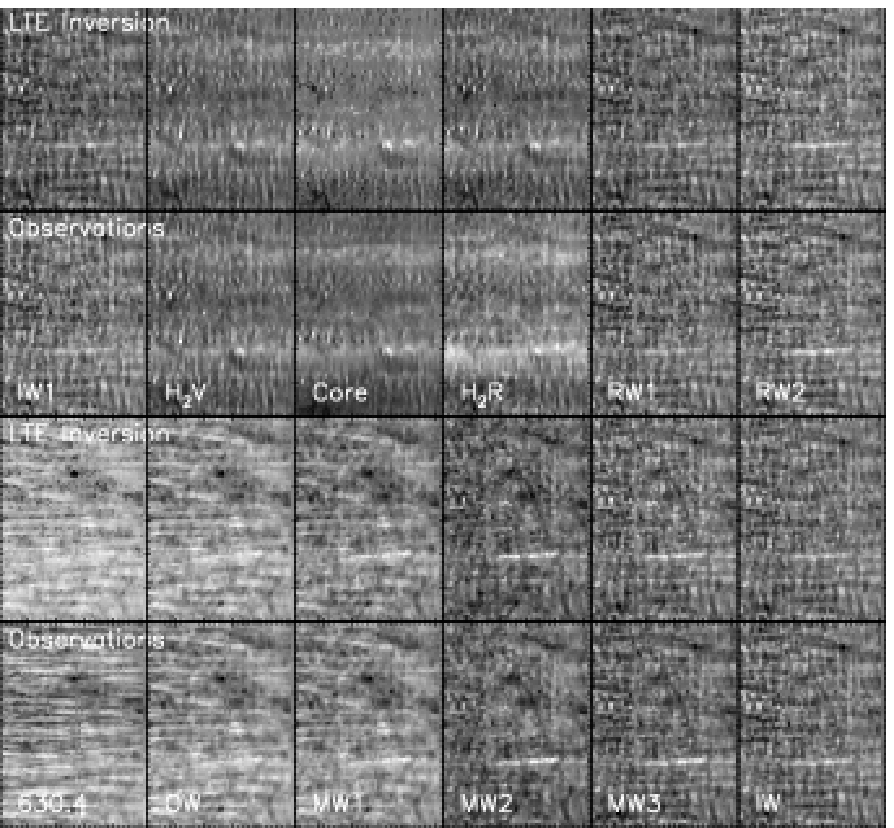}}$ $\\$ $\\
\end{minipage}}\hspace*{.25cm}
\begin{minipage}{5.cm}
\caption{{\em Bottom panel}: spatio-temporal maps of wavelengths bands in the observed spectra ({\em odd rows}) and in the best-fit spectra ({\em even rows}) of the time series. The {\em lower two rows} show {\em from left to right} the continuum intensity at 630.4 nm, OW, MW1, MW2, MW3, and IW. The {\em upper two rows} show IW1, H$_{\rm 2V}$, core intensity, H$_{\rm 2R}$, RW1, and RW2. {\em Top panel}: corresponding temperature maps at several optical depths from $-0.5$ to $-5.5$ in $\log \tau$ (denoted at the {\em top} in each subpanel). The minimum and maximum temperatures are denoted at the left and right end of the grey bar, respectively. The three values on each side in the lower row correspond to the layers of $\log \tau = -0.5/-1.5/-2.5$ ({\em from left to right}), those in the upper row to $\log \tau = -3.5/-4.5/-5.5$.\label{fig_spatiotemp}}
\end{minipage}
\end{figure*}

The accuracy in the reproduction of the observed profiles in the QS is visualized in more detail in Fig.~\ref{fig5} that shows the evolution of individual spectra during about three minutes in the time-series. In that time span, one wave front passes through the Ca line formation range, culminating in strong emission of the H$_{\rm 2V}$ peak when reaching the line core. The intensity near the line core rises gradually (profiles Nos.~23 and 24), yielding a profile with an elevated line wing close to the line core and a narrow absorption core without emission peaks in profile No.~25 ({\em leftmost middle panel} of Fig.~\ref{fig5}) that gives way to one with a strong H$_{\rm 2V}$ emission peak 20\,--\,40\,secs later at profile No.~27 ({\em rightmost middle panel} of Fig.~\ref{fig5}). The emission in the line core subsides again after the passage of the shock front (profiles Nos.~28 to 30), whereas at the end in the line wing already the next wave front is indicated (compare profiles Nos.~24 and 30). The individual profiles clearly demonstrate that the H$_{\rm 2R}$ emission peak is not reproduced by the fit, but the amplitude of H$_{\rm 2V}$ is matched well, e.g., in profile Nos.~26 to 29.
\subsection{Example spectra in active region off disc centre\label{ar_spec_exam}}
Figure \ref{spat_spec_ar} shows observed and best-fit spectra for three spatial cuts through one of the three active region maps in the same layout as in the {\em left panel} of Fig.~\ref{slit_spec}. The best-fit spectra differ more from the observed spectra than for the QS data on disc centre. No LOS velocities were included in the fit because for the line core of \ion{Ca}{ii} H it is difficult to determine any suitable value directly from the spectra. The profiles in the umbra show only a single emission peak  (where a fit of a Gaussian could be used), those in the penumbra usually double reversals (where the location of the central absorption could be used), but also mixtures of these two types or profiles with a plateau of emission appear. A determination of LOS velocities to be attributed to the Ca line core would thus require a (predetermined) choice of the method to derive the velocity because it is impossible to use a unique method over the complete FOV. However, because of the small impact of LOS velocities up to the sound speed on the spectra, neglecting the post-facto inclusion of velocities should not have influenced the retrieved temperature stratifications.
\begin{figure*}
\fbox{\begin{minipage}{17.6cm}Temperature\\
\resizebox{17.6cm}{!}{\hspace*{1.cm}\includegraphics{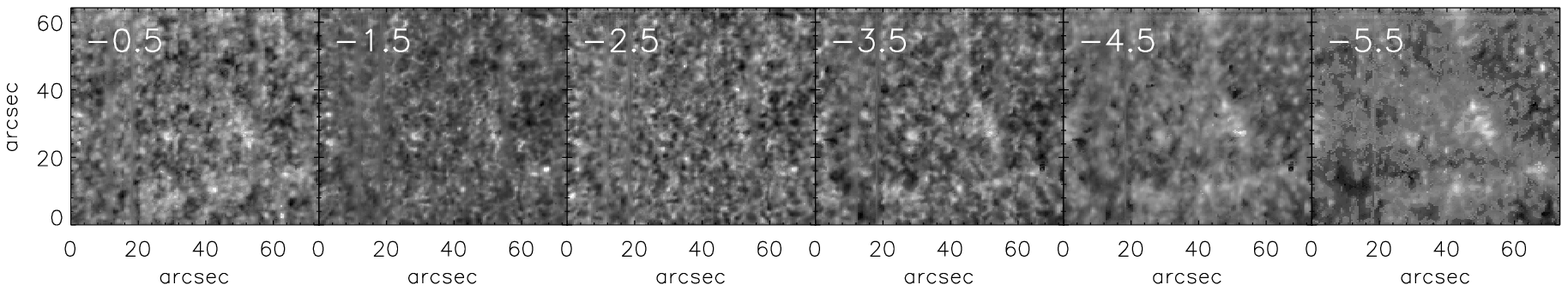}}$ $\\$ $\\$ $\\
\centerline{\hspace*{.25cm}\resizebox{8.cm}{!}{\includegraphics{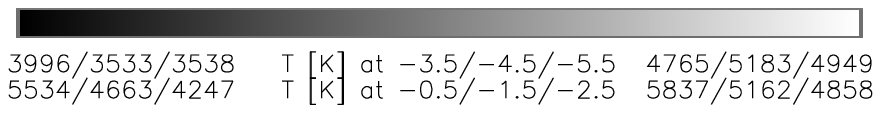}}}\end{minipage}}
\fbox{\begin{minipage}{17.6cm}
Wavelength bands\\
\centerline{\resizebox{17.6cm}{!}{\hspace*{1.cm}\includegraphics{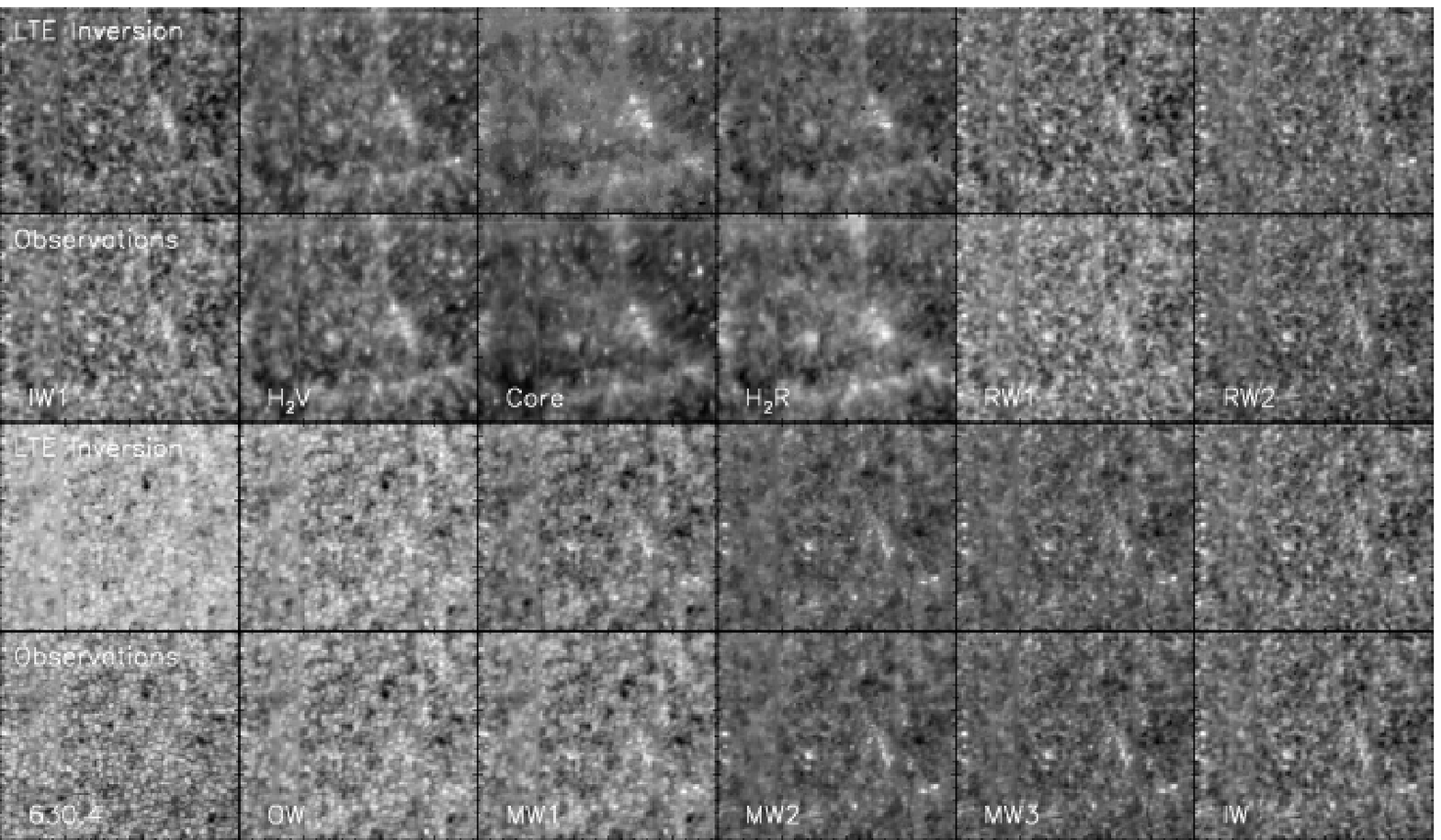}}}$ $\\$ $\\
\end{minipage}}
\caption{{\em Bottom panel}: 2D maps of wavelengths bands in the observed spectra ({\em odd rows}) and in the best-fit spectra ({\em even rows}) of the large-area scan on disc centre. The {\em lower two rows} show {\em from left to right} the continuum intensity at 630.4 nm, OW, MW1, MW2, MW3, and IW. The {\em upper two rows} show IW1, H$_{\rm 2V}$, the line-core intensity, H$_{\rm 2R}$, RW1, and RW2. {\em Top panel}: corresponding temperature maps at several optical depths from $-0.5$ to $-5.5$ in $\log \tau$ (denoted  {\em at the top} in each subpanel). See Fig.~\ref{fig_spatiotemp} for additional information on the grey bar.\label{fig_spatiospatio}}
\end{figure*}

Nevertheless, a clear deviation between best-fit and observed profiles is seen in the very line core of \ion{Ca}{ii} H. All best-fit profiles outside the umbra are characterized by a central absorption core that is neither seen in the observed AR spectra nor the QS data. This fact is better visible in the individual spectra of Fig.~\ref{fig_ar_single} that correspond to a position in the umbra, penumbra, a bright point (BP) in a plage area, and a ``QS'' location in the AR. Except for the umbral profile, all best-fit profiles have a dark central absorption core. This absorption core presumably results from the requirement to have a decrease of the temperature in the uppermost layers to obtain ``reasonable'' spectra \citep[imposed by the LTE assumption, cf.~also][]{rezaei+etal2008}. It shows up much more prominent for the AR LTE spectra because there a relatively strong chromospheric temperature rise is required to produce the strong emission core at first, on whose top the (small) absorption core then is superposed. In the QS spectra, the strong emission core is absent. Another difference between Figs.~\ref{spat_spec} and \ref{spat_spec_ar} is the temperature range in the AR required for displaying the temperature stratifications in the {\em third column} of Fig.~\ref{spat_spec_ar}. The variations of temperature are so much larger than in QS that the display range had to be significantly increased, even while the display now is on a logarithmic scale. 

The creation of the archive spectra was done in a semi-automatic way without explicitly taking into account differences between spectra in QS and AR. Figure \ref{fig_ar_single} demonstrates that the archive by chance also covered the range of profile shapes that are observed in ARs. 
\begin{figure*}
\begin{minipage}{17.6cm}
\centerline{\resizebox{17.6cm}{!}{\hspace*{.5cm}\includegraphics{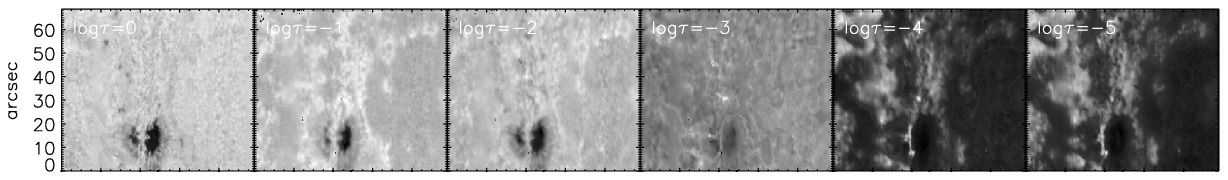}}}
\centerline{\resizebox{17.6cm}{!}{\hspace*{.5cm}\includegraphics{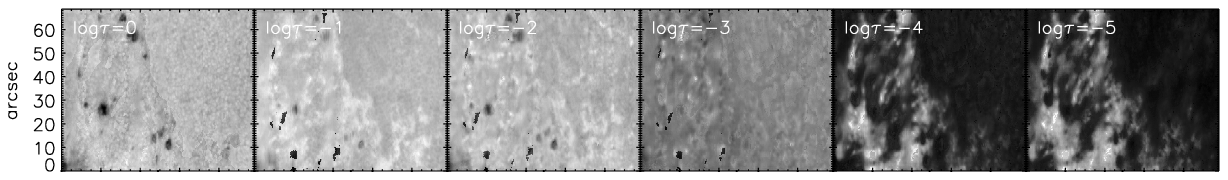}}}
\centerline{\resizebox{17.6cm}{!}{\hspace*{.5cm}\includegraphics{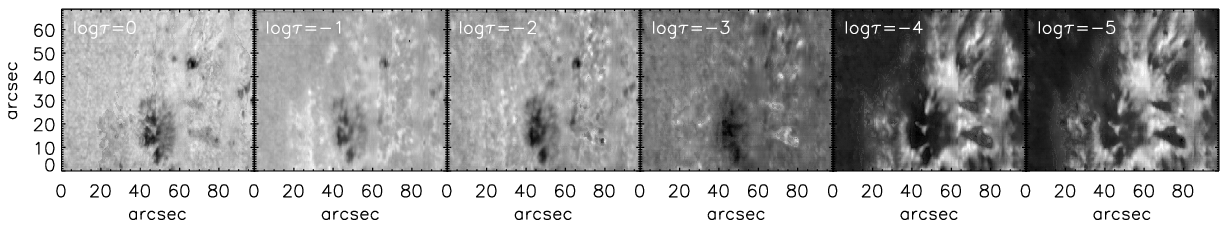}}}$ $\\$ $\\
\centerline{\hspace*{.25cm}\resizebox{8.cm}{!}{\includegraphics{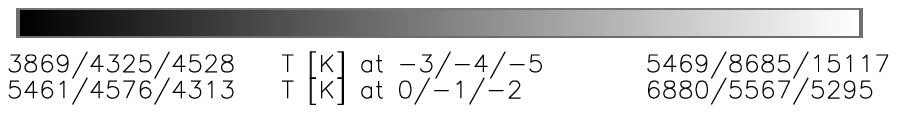}}}
\end{minipage}
\caption{Temperature maps for the active region scans. Note that the layers of $\log\tau$ slightly differ (0 to -5 in $\log \tau$). The temperature values of the grey bar refer to the {\em middle row} as reference, the range in the other two scans is of comparable order.\label{temp_active}}
\end{figure*}
\subsection{2D maps in quiet Sun}
The {\em lower panel} of Fig.~\ref{fig_spatiotemp} shows 2D spatio-temporal maps of the intensity in selected wavelength bands as final cross-check of the performance across the complete FOV and the full spectral range \citep[compare to, e.g.,][their Figs.~2 and 3]{cauzzi+etal2008a}. The wavelength bands are described in more detail in \citet[][]{rezaei+etal2007} and BE08, and are marked in the {\em upper left panel} of Fig.~\ref{fig5}. They pass from the outer line wing (OW) in the blue through the line core towards the red line wing (RW), avoiding the line blends. The only differences between observed and inverted spectra that can be discerned by eye are in the map of the continuum intensity at 630.4\,nm and in the H$_{\rm 2R}$ map. For the former, the dynamic range of the observed continuum intensity at 630.4\,nm is not fully matched and the spatial pattern in the inversion spectra is more similar to that in the OW  than to the observed continuum intensity itself. This presumably is related to the small influence of the continuum intensity in the iterative improvement of the fit because it provides only one wavelength, and hence one optical depth point to the modification of the temperature stratification (Sect.~\ref{mod_temp}). For H$_{\rm 2R}$, the observations show a higher intensity than the fit for the network regions at $y=10 - 20^{\prime\prime}$ and around 50$^{\prime\prime}$.

The {\em upper panel} of Fig.~\ref{fig_spatiotemp} shows temperature maps at several optical depth values for the inversion of the time-series, for a visualization of the relation between observed spectra and the corresponding temperatures. The LTE assumption maps the intensity at some wavelength to the temperature at some optical depth, but because the radiative transfer through the complete depth stratification was considered, the approach exceeds the direct conversion from intensity at one wavelength to temperatures through the Planck function used in, e.g., \citet{cauzzi+etal2009} or BE12. During the iteration of the best-fit stratification from the LTE archive the code also  always modifies the temperature over some range in $\log \tau$ such that the profile after the integration of the radiative transfer equation matches the observed spectrum. This also breaks up the one-to-one correlation between a given wavelength and a single layer in $\log \tau$. Comparing the intensity and temperature maps in Fig.~\ref{fig_spatiotemp}, one can identify some features that appear modified in the spectra and the temperature stratification. One case is a large-scale $\subset$-shaped darkening around ($t=40$\,min, $y=25^{\prime\prime}$) that appears only prominently in the line-core map of the observations and weakly in H$_{\rm 2V}$ and H$_{\rm 2R}$, but can be seen in the temperature maps for all layers above $\log\tau = -4.5$.

Figure \ref{fig_spatiospatio} shows the intensity maps in the wavelength bands in the large-area scan in QS on disc centre. For the comparison between observed and best-fit spectra in Fig.~\ref{fig_spatiospatio}, the match seems to be even better than for the spectra of the time-series on disc centre shown in Fig.~\ref{fig_spatiotemp}. The increasing spatial extent of the emission near the locations of photospheric network fields in wavelength bands near the line core is clearly seen in both observed and best-fit spectra. Wavelength bands that form lower are also well matched. The largest deviations can be seen in the map of the H$_{\rm 2R}$ emission peak ({\em fourth column} in the {\em upper two rows}), the RW1 map ({\em fifth column} in the {\em upper two rows}), and the continuum intensity at 630.4\,nm. The latter two maps show some intensity offset and contrast difference between observations and best-fit spectra. 

The temperature maps in the {\em upper panel} of Fig.~\ref{fig_spatiospatio} show the extension of the emission around the locations of concentrated photospheric magnetic flux for $\log \tau$ smaller than $-4.5$. Isolated temperature enhancements (e.g., at $x,y \sim 65^{\prime\prime}, 50^{\prime\prime}$) seem to be more pronounced and localized than the corresponding brightenings in intensity in H$_{\rm 2V}$ or the line-core map. The magnetic locations are not discernible in the temperature map at $\log\tau = -1.5$. The lowest layer displayed at $\log \tau = -0.5$ already  clearly forms above the continuum layers and corresponds roughly to the intensity map of MW2. 
\begin{figure*}
\sidecaption
\begin{minipage}{12cm}
\resizebox{12.cm}{!}{\hspace*{.5cm}\includegraphics{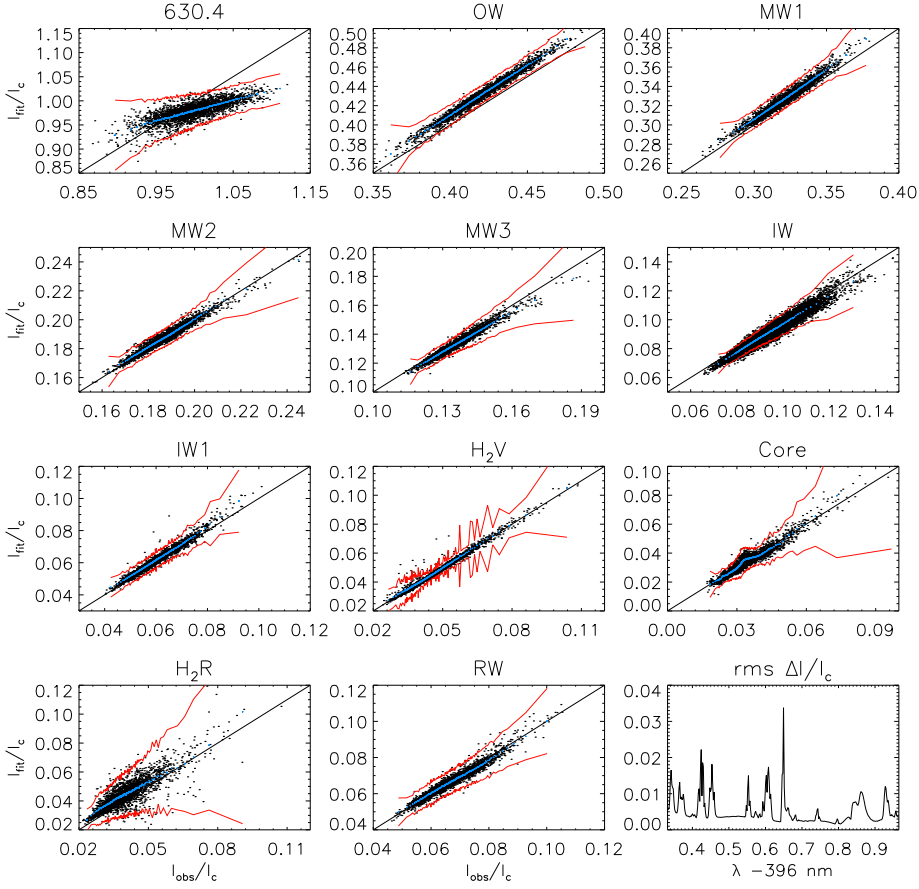}}
\end{minipage}
\caption{Scatter plots of observed and best-fit intensities ({\em black dots}) of the large-area scan on disc centre. The {\em blue dots} denote binned values whose 3-$\sigma$ variation is denoted by the {\em red lines}. The {\em solid black line} denotes a one-to-one correlation. {\em Bottom right panel}: standard deviation of the difference between observed and best-fit intensity as function of wavelength.\label{temp_large_scatterplot}}
\end{figure*}
\subsection{2D maps in active region}
In the temperature maps of the AR (Fig.~\ref{temp_active}) several peculiarities stand out. The map shown in the {\em bottom row} exhibits two dark filaments for $\log\tau < -4$ at $x\sim 70^{\prime\prime}$ and $y\sim 15^{\prime\prime}$ and $30^{\prime\prime}$, respectively. The lower, larger one has a co-spatial filament in \ion{He}{i} spectra at 1083\,nm, and was identified with a transient siphon flow by \citet{bethge+etal2012}. Several similar dark filaments that originate or terminate in individual pores can be identified in the scan shown in the {\em middle row}. An investigation of the related Doppler shifts will allow one to determine if they are also related to similar siphon flows \citep[cf.][]{uitenbroek+etal2006,beck+etal2010}. The AR map in the {\em top row} shows two interesting emission features, or equivalently locations of strong temperature enhancements. The first is an isolated bright point (BP) at $(x,y) \sim (45^{\prime\prime},30^{\prime\prime})$. This BP has about the highest temperature of the full FOV at $\log\tau= -3$, and still stands out in the map at $\log\tau= -4$. The Ca spectra at this location are similar to the BP spectra shown in  Fig.~10 of \citet{beck+etal2005b}. The second large-scale region of temperature enhancements is located just between the large and small sunspot in the {\em top row} of Fig.~\ref{temp_active} at $(x,y) \sim (40^{\prime\prime},10^{\prime\prime}-20^{\prime\prime})$. The increase of emission between the two sunspots at $\log\tau= -3$ to $-5$ could indicate the presence of a current sheet that separates their field lines, similar to the current sheet discussed in \citet{tritschler+etal2008} for \ion{Ca}{ii} IR observations \citep[see also][]{goodman+judge2012}.
\section{Discussion \label{discussion}}
\subsection{Quality in reproduction of observed spectra \label{qual_fit}}
The comparison of the 2D intensity maps of observed and best-fit spectra showed some differences between the two. To quantify the resulting mismatch, we made scatter plots of the observed and best-fit intensities in the wavelength bands for the large-area scan on disc centre (Fig.~\ref{temp_large_scatterplot}). The graphs confirm that the intensity in the continuum at 630.4\,nm and in the H$_{\rm 2R}$ emission peak deviate between observations and best-fit spectra: for the former, the slope differs from unity, while for the latter the scatter is significantly larger than in, e.g., H$_{\rm 2V}$. For all other wavelength bands the 3-$\sigma$ interval of scatter ({\em red lines}) covers the line of one-to-one correlation. The {\em bottom right panel} of Fig.~\ref{temp_large_scatterplot} shows the standard deviation of the difference between the observed and best-fit intensities as a function of wavelength. Besides from the location of the line blends, the rms deviation is about 0.5\,\% of $I_c$ in the line wing, and increases to about 1\,\% of $I_c$ for wavelengths in the line core from 396.82 to 396.88\,nm. The rms at the location of the H$_{\rm 2R}$ emission peak is slightly enhanced relative to rest of the line core. 
\begin{figure}
\resizebox{8.8cm}{!}{\hspace*{1.cm}\includegraphics{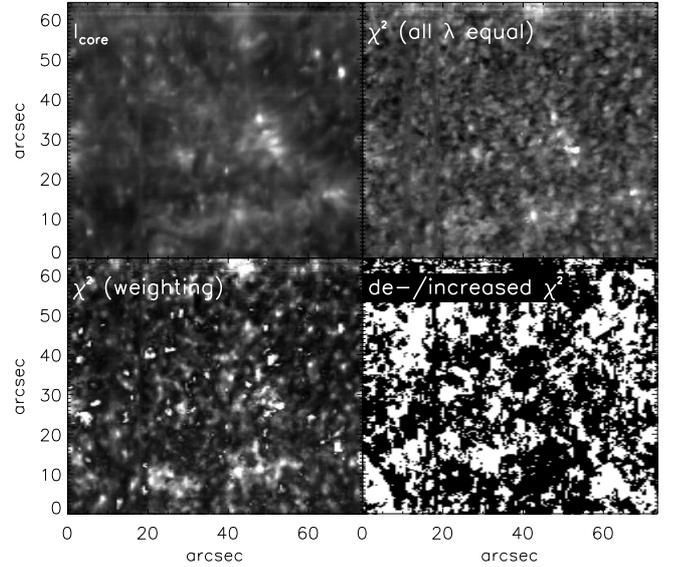}}$ $\\$ $\\
\caption{Spatial variation of the $\chi^2$ in the large-area QS scan. {\em Top row, left}: line-core intensity as indicator of the solar spatial structuring. {\em Top right}: $\chi^2$ with equal weight for all wavelengths. {\em Bottom left}: $\chi^2$ with enhanced weighting for the line core. {\em Bottom right}: decrease/increase of $\chi^2$ ({\em white/black}) by the iterative modification of the fit when considering enhanced weighting in the line core.\label{chi_2d}}
\end{figure}

The spatial variation in the reproduction of the observed spectra is displayed in Fig.~\ref{chi_2d}. In the map of $\chi^2$ with equal weighting for all wavelengths, i.e., $w(\lambda) \equiv\,1$, that measures the quality of the fit to the full line profile ({\em top right panel}), the centres of the network regions stand out, but only slightly. The cellular granular-sized pattern in the $\chi^2$-map corresponds to spatial structures seen in wavelengths forming at and below the MW and indicates misfits in the lower atmosphere. Calculating the $\chi^2$ with the enhanced weighting for the line core as used in the retrieval of the best-fit archive profile ({\em bottom left panel}) yields a similar spatial variation of the fit quality, with the network locations now standing out more prominently and a more homogeneous quality throughout the inter-network regions. The marker that shows how the $\chi^2$ with the enhanced weighting for the line core was modified by the iterative improvement of the fit is displayed at the {\em lower right}. It turns out that on about 60\,\% of the FOV, the iterative modification of the full temperature stratification has worsened the initial fit to the very line core obtained using only the best-fit archive profile. Only in the inter-network regions with usually less complex profile shapes, the iteration has improved not only the fit to the line wing, but also to the line core even with the strongly enhanced contribution of the line core. For the AR data, in nearly all cases ($>90\,\%$) the fit to the line core was worsened in the iterative modification. We note that this is a slight worsening relative to the best-fit archive profile, with still an acceptable fit to the line shape even for most of the AR spectra (e.g., Fig.~\ref{fig_ar_single}).
\subsection{Statistics on the archive usage}
\begin{figure}
\resizebox{8.8cm}{!}{\includegraphics{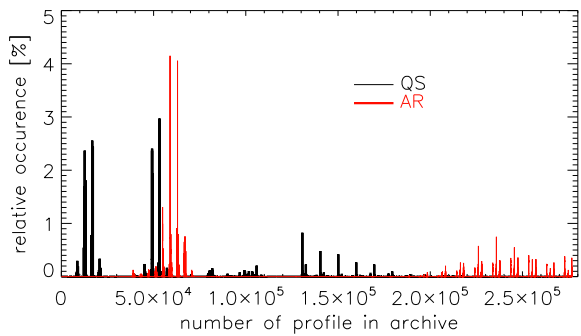}}
\caption{Relative occurrence of a given archive profile in the inversion. {\em Thick black}: QS map. {\em Thin red}: AR map.\label{archive_stat}}
\end{figure}

The size of the pre-calculated spectral archive partly limits the usefulness of the two-step inversion approach described here. At the spectral sampling of the POLIS data of 1.92\,pm per pixel and the total wavelength range covered, the full archive of about 300.000 spectra could be easily kept in the memory of a simple standard desktop machine because its size was only about 350\,MB. Figure \ref{archive_stat} shows the statistics on how often which archive profile was actually used in the determination of the best-fit archive profile. The most frequently used archive profiles appear with about 3\,--\,4\,\% relative frequency. Some areas in the archive were only relevant in either the QS or the AR data, e.g., the archive profiles with numbers above 200.000 only appear for the AR spectra. These profiles correspond to a large positive offset $T_{diff}$ in temperature in their synthesis. This global offset in temperature between the QS and AR data is caused by both the change in the heliocentric angle, where in the AR data the LOS reaches a given optical depth at a different geometrical height in comparison to the QS data, and the presence of magnetic fields throughout nearly all of the FOV in the AR that is known to lead to increased emission in Ca spectra \citep{rezaei+etal2007,rezaei2008}.

It turns out that the actual usage of the archive is much more sparse than the plot of Fig.~\ref{archive_stat} suggests. In total, only about 5700 different archive profiles ($\sim 2\,\%$ of the full archive) were picked in the inversion of either QS or AR data ($\sim$80.000 observed spectra). The acceptable quality of the fits in both QS and AR implies that an exchange of the LTE archive for about 10.000 profiles calculated in NLTE conditions could already suffice to provide a reasonable NLTE fit to observed chromospheric \ion{Ca}{ii} H profiles, and hence as well to similar spectra in other chromospheric lines. The full dynamical range of the temperature variations seems to be more critical in the creation of the spectral archive than the density of the sampling inside the parameter space of temperature.
\subsection{Outlook}
Our findings can be extended to what will be needed for the future analysis of spectra from, e.g., the planned Solar-C mission, regardless of a decision for the \ion{Ca}{ii} H and K or the \ion{Mg}{ii} h and k lines \citep{katsukawa+etal2011,beluzzi+trujillobueno2012} because of their similar behaviour, or data from the planned BLue Imaging Solar Spectrometer \citep[BLISS,][in press]{puschmann+etal2012a} for the new 1.5-m GREGOR telescope \citep[e.g.,][in press]{schmidt+etal2012}. The BLISS data will be accompanied by spectropolarimetric observations at, e.g., 630\,nm with the GREGOR Fabry-Per\'ot Interferometer \citep[][in press]{puschmann+etal2012}, which will provide the possibility to simultaneously measure photospheric magnetic fields as in the case of POLIS.

For deriving the thermodynamic properties of the solar chromosphere from observed spectra, direct analysis tools will be few and very limited. The best approach seems to consist in a ``controlled'' NLTE forward modeling, i.e., a creation of NLTE spectra for what can be considered realistic solar atmosphere models. These could be derived from both analytical 2D solutions of solar surface structures \citep{uitenbroek2011} or synthetic spectra from numerical simulations \citep{leenaarts+etal2009,leenaarts+etal2010,wedemeyer+carlsson2011}. The lateral radiative transport is more important for chromospheric than for photospheric lines. A creation of an archive from 3D numerical simulations might thus be anyway required for a consistent spectral synthesis \citep[cf.][]{uitenbroek+criscuoli2011,sheminova2012,leenaarts+etal2012}. Such an archive then also allows one to improve the inversion procedure beyond the common single-pixel approach in which every pixel is inverted individually neglecting its surroundings. Especially for chromospheric lines, a simultaneous fit to a small-scale area of, e.g., about $1^{\prime\prime}\times1^{\prime\prime}$ ($3\times3$ pixels at the POLIS spatial sampling) can be crucial to obtain consistent results. An NLTE archive from 3D numerical simulation can directly provide suited spectra on a 2D area including a central pixel and its surroundings. There will be a need for additional modification of initial profiles and the related thermodynamical variables to match the observed spectra, similar to the two-stage approach taken here, because the combined parameter space of the uncorrelated temperature and velocity variations cannot be covered by any reasonably sized archive.

The creation of a base of NLTE spectra seems to be a promising approach for both the Ca and Mg lines, with the advantage that such spectra can already be created and tested against simulations and/or theoretical models \citep[cf.][]{delacruz+etal2012} well in advance of the real data to be obtained in a few years. 

The results of the retrieved temperature stratification in QS and AR will be discussed in the subsequent paper of this series.
\section{Conclusion\label{conclusion}}
A pre-calculated archive of about 300.000 spectra is found to be more than sufficient to fit the line cores of observed \ion{Ca}{ii} H profiles in quiet Sun and active regions satisfactorily regardless of the huge range of possible shapes, whereas the line wing has to be iteratively modified, because line core and line wing vary independently of each other. The advantage of the two-step inversion approach is that an exchange of the archive for a suited data base of spectra calculated in NLTE suffices to provide a first-order NLTE inversion method. 

We have to leave it up to the reader to decide if we met the requirement given in the quotation from \citet{cram1977} at the very beginning to provide a ``careful study''. By using the LTE assumption, we are, however, perfectly sure that we fulfill his labeling of using a ``crude'' method.  
\begin{acknowledgements}
The VTT is operated by the Kiepenheuer-Institut f\"ur Sonnenphysik (KIS) at the
Spanish Observatorio del Teide of the Instituto de Astrof\'{\i}sica de Canarias (IAC). The POLIS instrument has been a joint development of the High Altitude Observatory (Boulder, USA) and the KIS. C.B.~acknowledges partial support by the Spanish Ministry of Science and Innovation through project AYA2010--18029 and JCI-2009-04504. R.R. acknowledges financial support by the DFG grant RE 3282/1-1. We thank M.~Collados, H.~Balthasar, J.~Staude, and H.~Socas-Navarro for their comments on the manuscript.
\end{acknowledgements}
\bibliographystyle{aa}
\bibliography{references_luis_mod_partIII}
\begin{appendix}
\section{Animation of inversion results\label{appa}}
  
\begin{figure*}
\centerline{\resizebox{17.6cm}{!}{\hspace*{.5cm}\includegraphics{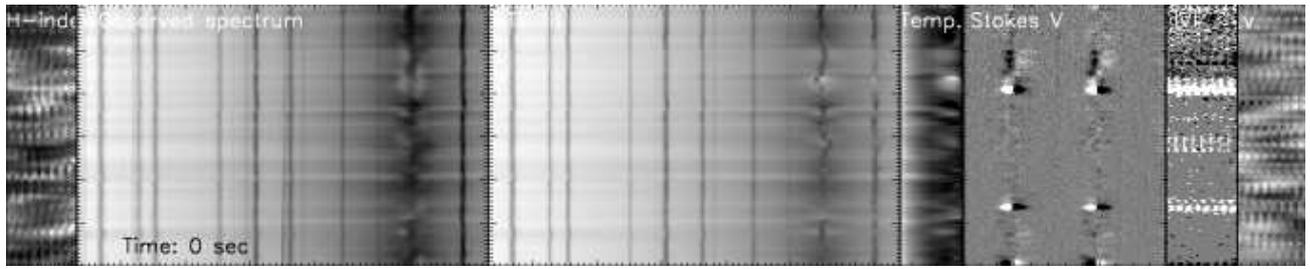}}}$ $\\
\caption{Still from an animation of the inversion results for the time-series of quiet Sun. {\em Leftmost panel}: 2D maps of the H-index for eight repetitions of scanning the same area of $2^{\prime\prime}\times 61^{\prime\prime}$. Time increases left to right. The scan step of the spectra shown in the other panels is always located at the very left border at 0$^{\prime\prime}$. {\em Second and third panel}: observed and best-fit Ca spectra. {\em Fourth panel}: temperature stratifications along the slit corresponding to the best-fit spectra. {\em Fifth panel}: co-spatial and simultaneous Stokes $V$ spectra at 630\,nm. {\em Sixth panel}: 2D maps of the polarity of the Stokes $V$ signal for eight repetitions of scanning the same area. {\em Seventh panel}: 2D maps of the LOS velocity of the \ion{Fe}{i} line at 630.15\,nm for eight repetitions of scanning the same area.  \label{app_fig}}
\end{figure*}
Figure \ref{app_fig} shows a still from an animation that displays the inversion results and some line characteristics from both the Ca and the 630\,nm spectra for the time-series of quiet Sun. There are three 2D maps: the H-index (integration of the line-core intensity of Ca) in the {\em first panel}, the polarity of the Stokes $V$ signal ($\pm 1$ for magnetic fields along/anti-parallel to the LOS and zero for locations without significant polarization signal) in the {\em sixth panel}, and the LOS velocity of the \ion{Fe}{i} line at 630.15\,nm in the {\em seventh panel}. For all these 2D maps, eight repetitions of scanning the same area of $2^{\prime\prime}\times 61^{\prime\prime}$ are shown, i.e., the temporal evolution inside the area over the course of about 170\,seconds is displayed. The spectra shown in the rest of the panels were always taken from the scan step that is located at 0$^{\prime\prime}$ in the 2D maps. The animation runs over the complete duration of the time-series of about one hour. 

In the 2D maps of the H-index, the lateral spatial motion of individual bright grains can be followed. The temporal evolution of the temperature stratifications shows the spatial (along the slit and in height) and temporal extent of wave trains and temperature perturbations. One prominent example of an upwards propagating temperature perturbation that subsequently affects the full line profile appears between $t = 378$ and $483$\,seconds at $y\sim 28^{\prime\prime}$, i.e., at about the middle of the FOV along the slit. Lateral motion of temperature perturbations along the slit can be seen for some events that happen close to locations with strong polarization signal. One such event shows up between $t = 693$ and $756$\,seconds, starting at the location of the persistent polarization signal at $y\sim 40^{\prime\prime}$ and moving downwards along the slit with time, while getting increasingly weaker on the way. The Stokes $V$ spectra and the polarity of the Stokes $V$ signal show that most of the polarization signals, and hence photospheric magnetic fields detected are rather stable throughout the time-series. With an integration time of less than 4 seconds per scan step, the polarimetric sensitivity in the time-series is naturally limited to strong polarization signals that correspond to the stable photospheric magnetic network. The polarity was erroneously labeled with ``$|V|$'' during the creation of the animation.
\end{appendix}
\end{document}